\documentclass[conference]{IEEEtran}
\IEEEoverridecommandlockouts
\usepackage{cite}
\usepackage{amsmath,amssymb,amsfonts}
\usepackage{algorithmic}
\usepackage{graphicx}
\usepackage{subcaption}
\usepackage{makecell}
\usepackage[american]{circuitikz}
\usepackage{textcomp}
\usepackage{xcolor}
\usepackage{todonotes}
\usepackage{balance}

\usepackage[all]{background}

\usepackage{stackengine}
\setstackEOL{\\}
\setstackgap{L}{\normalbaselineskip}
\SetBgContents{\color{gray}{\tiny \Longstack{PREPRINT - accepted by 21st IEEE Interregional NEWCAS Conference 2023 (NEWCAS 2023)}}}
\SetBgPosition{3.5cm,1cm}
\SetBgOpacity{1.0}
\SetBgAngle{0}
\SetBgScale{1.8}

\def\BibTeX{{\rm B\kern-.05em{\sc i\kern-.025em b}\kern-.08em
    T\kern-.1667em\lower.7ex\hbox{E}\kern-.125emX}}

\def\@IEEEfigurecaptionsepspace{\vskip\abovecaptionskip\relax}%
\def\@IEEEtablecaptionsepspace{\vskip\abovecaptionskip\relax}%
    
\begin{document}

\title{Finite State Automata Design using 1T1R ReRAM~Crossbar\vspace{-5mm}}

\author{\IEEEauthorblockN{Simranjeet Singh\IEEEauthorrefmark{1}\IEEEauthorrefmark{5}, Omar Ghazal\IEEEauthorrefmark{2}, Chandan Kumar Jha\IEEEauthorrefmark{3}, Vikas Rana\IEEEauthorrefmark{5},
Rolf Drechsler\IEEEauthorrefmark{3}\IEEEauthorrefmark{4}, \\
Rishad Shafik\IEEEauthorrefmark{2}, 
Alex Yakovlev\IEEEauthorrefmark{2},
Sachin Patkar\IEEEauthorrefmark{1},
Farhad Merchant\IEEEauthorrefmark{2} \IEEEauthorblockA{\IEEEauthorrefmark{1}Indian Institute of Technology Bombay, India, \IEEEauthorrefmark{2}Newcastle University, UK, \IEEEauthorrefmark{3}University of Bremen, Germany, \\ \IEEEauthorrefmark{4}DFKI GmbH, Germany, \IEEEauthorrefmark{5}Forschungszentrum Jülich GmbH, Germany}}
\{simranjeet, patkar\}@ee.iitb.ac.in, \{si.singh, v.rana\}@fz-juelich.de, \{chajha,drechsler\}@uni-bremen.de, \\ \{O.G.G.Awf2, rishad.shafik, alex.yakovlev, farhad.merchant\}@newcastle.ac.uk \vspace{-7mm}}
\maketitle

\begin{abstract}
Data movement costs constitute a significant bottleneck in modern machine learning (ML) systems. When combined with the computational complexity of algorithms, such as neural networks, designing hardware accelerators with low energy footprint remains challenging.
Finite state automata (FSA) constitute a type of computation model used as a low-complexity learning unit in ML systems. The implementation of FSA consists of a number of memory states. However, FSA can be in one of the states at a given time. It switches to another state based on the present state and input to the FSA. Due to its natural synergy with memory, it is a promising candidate for in-memory computing for reduced data movement costs. 
This work focuses on a novel FSA implementation using resistive RAM (ReRAM) for state storage in series with a CMOS transistor for biasing controls. We propose using multi-level ReRAM technology capable of transitioning between states depending on bias pulse amplitude and duration. We use an asynchronous control circuit for writing each ReRAM-transistor cell for the on-demand switching of the FSA. We investigate the impact of the device-to-device and cycle-to-cycle variations on the cell and show that FSA transitions can be seamlessly achieved without degradation of performance. Through extensive experimental evaluation, we demonstrate the implementation of FSA on 1T1R ReRAM crossbar. 

\end{abstract}

 \vspace{-3mm}
\section{Introduction}
\label{sec:intro}
In-memory computing (IMC) using memristive devices has become popular in elevating the von Neumann bottleneck by storing and processing data in memory, especially for machine learning (ML) applications~\cite{Giacomo2021,Yu2018,jha2022,Froehlich2022, Staudigl2022}.
Memristive devices connected in a crossbar structure allow the program to run in parallel, making the IMC comparable with conventional computing in terms of energy efficiency and performance~\cite{Sebastian2020}. Memristive devices, such as resistive random access memory (ReRAM)~\cite{waser2009}, can be configured as multi-level cell~\cite{Siemon2019}, where the device has multiple intermediate states between low resistive (LRS) and high resistive state (HRS). Still, modern ML workloads require massive storage resources and parallelism to accelerate. However, finite state automata (FSA) capture the real-world constraint of finite memory to implement learning applications~\cite{rabin1959}. The multi-level behavior of ReRAM can be used to design FSA in memory.

\begin{figure}[!t]
    \centering
    \includegraphics[width=0.9\linewidth]{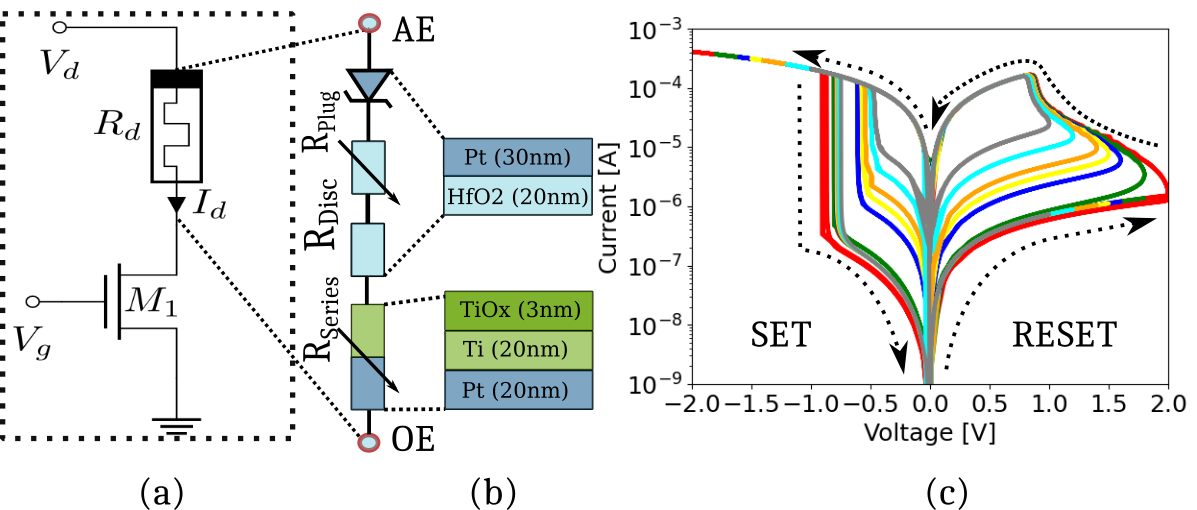}
    \caption{1T1R cell for FSA (a) cell structure, (b) device material stack, and (c) multi-level characteristic of a device.}
    \label{fig:1T1R}
\end{figure}

FSA is an abstract machine that can be at exactly one of the finite number of states at any given time. FSA changes its state from one to another, called a transition, when a specific event occurs~\cite{Rezvanian2019}. FSA has  a significant edge on applications that require low-latency or real-time processing, such as automated verification~\cite{luo2019}, autonomous vehicles~\cite{Zhang2017}, and tsetlin machine~\cite{Granmo2018}. The energy efficiency, high density, and IMC properties of ReRAM devices make them suitable for FSA implementation~\cite{wheeldon2020learning}~\cite{Halawani2019}. Multiple states of ReRAM, between LRS and HRS, can be mapped to the states of FSA. An FSA cell can be represented as a single ReRAM device with a CMOS transistor in series (1T1R), as shown in Fig.~\ref{fig:1T1R}(a). The material stack of the memristive cell is shown in Fig.~\ref{fig:1T1R}(b), along with the I-V characteristics of the 1T1R cell in Fig.~\ref{fig:1T1R}(c). Fig.~\ref{fig:1T1R}(c) shows the multiple states in SET (switching to LRS) and RESET (switching to HRS) states, which have been utilized in this study to implement the FSA on ReRAM devices~\cite{Aziza2021}.

However, transitions of FSA from one state to another with accurate detection of the current FSA state under device variations remains challenging. In this paper, we propose an architecture to implement the FSA on the ReRAM crossbar for IMC. We show in the proposed architectures how a single 1T1R cell can be used to design a six-state FSA. Next, we evaluate the proposed architecture in terms of energy efficiency and performance. Moreover, we assess the architecture under device variations such as device-to-device (D2D) and cycle-to-cycle (C2C)~\cite{Bengel2020}. To summarize the main contributions:

\begin{itemize}[]
\item The integration of 1T1R ReRAM technology into FSA design by utilizing the multi-level behavior of ReRAM. The gradual RESET method has been utilized to achieve the multi-level behavior.
\item Investigation of the impact of D2D and C2C variations on state transitions and detection.
\item Extensive evaluation of the efficiency of the proposed architecture in terms of energy efficiency and latency.
\end{itemize}

The remainder of the paper is organized as follows: Section~\ref{sec:arch} presents the proposed architecture to design FSA on ReRAM. Section~\ref{sec:Experimental_results} presents experimental results and validation for the design. Finally, Section~\ref{sec:conclusions} concludes the~paper.
\vspace{-1mm}
\section{Proposed Architecture}
\label{sec:arch}
\begin{figure}[!t]
    \centering
    \includegraphics[width=0.9\linewidth]{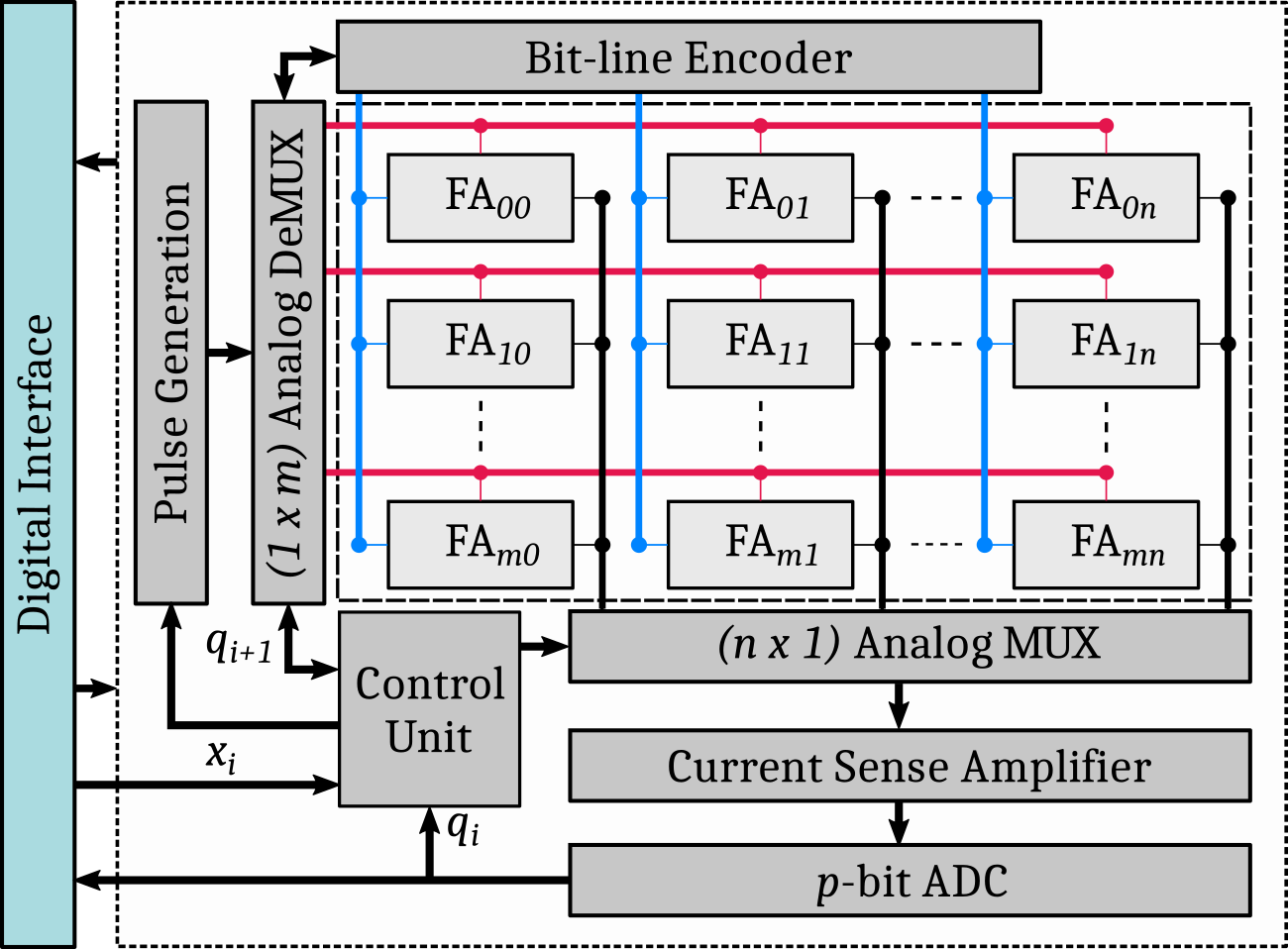}
    \caption{Architecture to train the FSA using 1T1R cell ($FA_{mn}$), where `\textit{m}' and `\textit{n}' represent rows and columns in a crossbar, respectively, and `\textit{p}' is the number of ADC bits, which is given as $\lceil \log_{2}(s) \rceil$, `\textit{s}' is the number of states in a FA~cell.\vspace{-7mm}}
    \label{fig:arch}
\end{figure}

This section discusses the architecture to implement the FSA on the 1T1R ReRAM crossbar, depicted in Fig.~\ref{fig:arch}. At the core of the proposed architecture, 1T1R cells connected in a crossbar structure called finite automaton (FA) are used. 

\subsection{FA using 1T1R cell}
The FA cell is constructed using one memristor and one NMOS transistor in series. The structure of a single FA cell is shown in Fig.~\ref{fig:1T1R}.The multi-level characteristics of the FA, which have been mapped to different states, are shown in the I-V characteristics displayed in Fig.~\ref{fig:1T1R}(c). In this study, the FA has seven states (`\textit{s}') from $S_0$ to $S_6$, starting from LRS to HRS. $S_0$ has a minimum resistance of around $7.8$K$\Omega$ and $S_6$ has maximum resistance of around $1.5$M$\Omega$. All other states are mapped into the intermediate values between $S_0$ to $S_6$. Each FA in the crossbar represents six states (excluding $S_0$), and they can be independently programmed. However, multiple FAs can be combined to run a complex application that needs more than six states. For this work, we will limit our study to the working of independent FA in the crossbar. The state transitions of FA are examined next.
\vspace{-1mm}
\subsection{State transitions in FA}
\vspace{-1mm}
An FA has a finite number of states, seven in this study, and it changes the state from one to another or the next state ($q_{i+1}$) based on the input ($i_i$) and current state ($q_i$) similar to a mealy machine. A pulse generation module in Fig.~\ref{fig:arch} can generate the different pulse widths of a fixed voltage amplitude. The control circuit selects the appropriate signal for the next state based on the present state and input. The parameter to switch the state from $S_0$ to any possible state is given in Table~\ref{tab:states}. Next, the analog demultiplexer (DeMUX) and bit-line encoder select a FA in the crossbar by applying an ON voltage to the NMOS transistor and the required pulse signal at the row of the crossbar. 

For the functional correctness of the FSA transition, it is important to identify the present state correctly. FSA can jump from the present state to any other possible state in FA. So, it is expected from the FSA that it should give the same current value for the same state transition. However, the gradual RESET method limits the transitions of the states only in the forward direction ($S_1 \rightarrow S_6$). In order to change the state which is less than the current state (backward direction), FA needs to switch to $S_0$ (intermediate state) before switching to the next desired state. Also, it is expected that states after switching can correctly be identified during forward or backward direction switching. Therefore, an intermediate state is added in every state transition in FA, which provides three main advantages; (a) switching to any state in FA, (b) state retention while looping in the same state, and (c) reducing the complexity of control circuitry. 
\begin{figure}[!]
\setlength\abovecaptionskip{-0.04\baselineskip}
\setlength\belowcaptionskip{-1.0\baselineskip}
   \centering
   \includegraphics[width=0.85\linewidth]{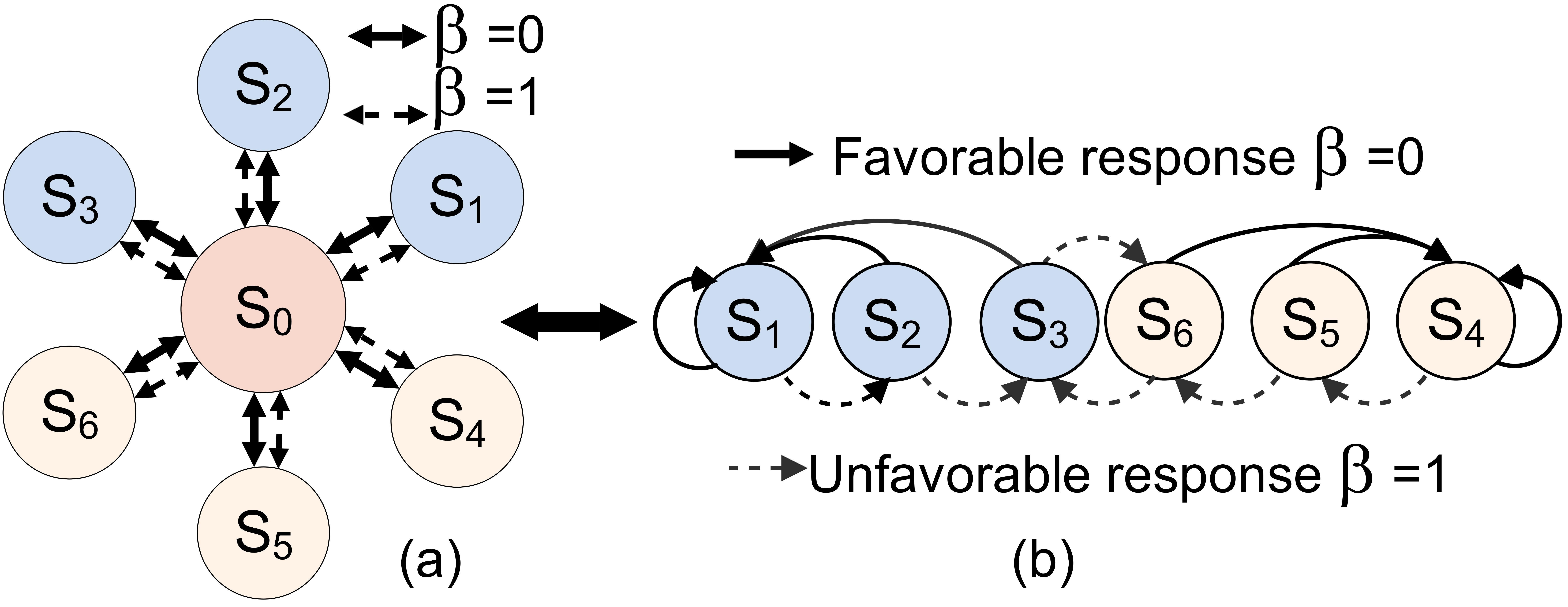}
    \caption{State transitions in a FA cell (a) and (b) shows the use of FSA as Krinsky learning automaton~\cite{Iraji2006}.\vspace{-3.2mm}}
   \label{fig:apps}
\end{figure}

The state transition graph of single FA is shown in Fig.~\ref{fig:apps}(a). It can be used for learning applications such as Krinsky automaton~\cite{Iraji2006}. The mapping of a Krinsky learning automaton has been shown in Fig.~\ref{fig:apps}(b). The unfavorable response ($\beta=1$) gradually moves toward the boundary states separating the two actions, behaving as binary states. In another response ($\beta=0$), $S_i$ switches to $S_{1}$ and $S_{4}$ for $(1\le i\le3)$ and $(4\le i\le6)$, respectively. The proposed architecture is highly flexible regarding its control circuit and switching characteristics. It provides the facility to transit from the current state to any next state via $S_0$ with an adaptive STG control unit. The proposed approach can accommodate FSM with more than 6 states by utilizing multiple 1T1R cells arranged to represent different states and encoded into binary form, offering flexibility for varying numbers of states.



\begin{table}[!b]
 \setlength\abovecaptionskip{0\baselineskip}
\setlength\belowcaptionskip{-1.5\baselineskip}
    \centering
    \caption{State transition of 1T1R cell}
     \label{tab:states}
\begin{tabular}{|c|c|c|c|}
    \hline
    \multicolumn{4}{|c|}{ \makecell{
    $\rm V_{fixed}$ = 1.8V,  $\rm V_{SET}$= $-2V$, $\rm V_{READ}$= $0.1V$\\
    }
    } \\
    \hline
         State &  $\rm P_{width}$ ($n$s) & $I_{d}$ ($\mu$A) & Resistance (K$\Omega$)\\
         \hline
        S0& 10$n$s at -2V& 12.8 & 7.8\\
        S1& 5$n$s & 12.6 & 8.0\\
        S2 & 10$n$s & 1.6 & 95.2\\
         S3 & 15$n$s & 0.56 & 196.1\\
        S4 & 30$n$s &0.3& 342.5\\
        S5 & 60$n$s &0.2& 588.2\\
       S6 & 150$n$s & 0.07& 1492.5\\
         \hline
    \end{tabular}

\end{table}
\vspace{-2mm}
\subsection{Peripherals to control FSA}
\vspace{-1mm}
Various peripherals around the crossbar are required to implement FSA on the ReRAM crossbar, as shown in Fig.~\ref{fig:arch}. The control unit decided the transitions in FA, which are functions of $x_i$ and $q_i$. 

\emph{Pulse generation module} generates the voltage pulses with a required duration for transitioning the state from one to another according to Table~\ref{tab:states}. Since state $S_0$ is an intermediate state, the small gap between $S_0$ and $S_1$ will not pose an issue in state estimation.

\emph{Multiplexer (MUX), demultiplexer (DeMUX), and bit-line encoder} are the selection peripherals. For a given ($m\times n$) size crossbar, ($1\times m$) sized DeMUX are attached to select a row of the crossbar to apply a pulse for transition. Bit-line encoder enables the transistor of selected FA. At column, ($n \times 1$) MUX is connected to read the state of a FA.

\emph{Current sense amplifier (CSA) and Analog-to-digital converter (ADC)} are sensing peripherals. The CSA converts the current to an amplified voltage, which is further used by the ADC to detect the current state of FA. A common CSA and ADC have been used in the proposed architecture, where a FA can be read in each cycle. A `\textit{p}' bit ADC is required to correctly detect the `\textit{s}' number of states in FA.

\emph{Control unit} contains the algorithms to switch the states of FA in sequence. The control unit manipulates the select lines of MUX and DeMUX to select an FA and generate the required pulses for transitions via the pulse generation module. It takes input from the digital interface ($x_i$) and ADC ($q_i$) and calculates the required control signal to switch the state to $q_{i+1}$; alternatively, $q_{i+1} (q_i,x_i)$, where $0<= |x_i| <=1$ and $S_1 <= q_i <= S_6$. \vspace{-2mm}

\section{Experimental Results}
\label{sec:Experimental_results}
This section evaluates the proposed methodology in terms of energy efficiency and area. First, we study the switching characteristics of a FA and the state transitions from one to another. Next, we look at the impact of D2D and C2C variations on state transitions.

\begin{figure}[!t]
    \centering
    \includegraphics[width=0.9\linewidth]{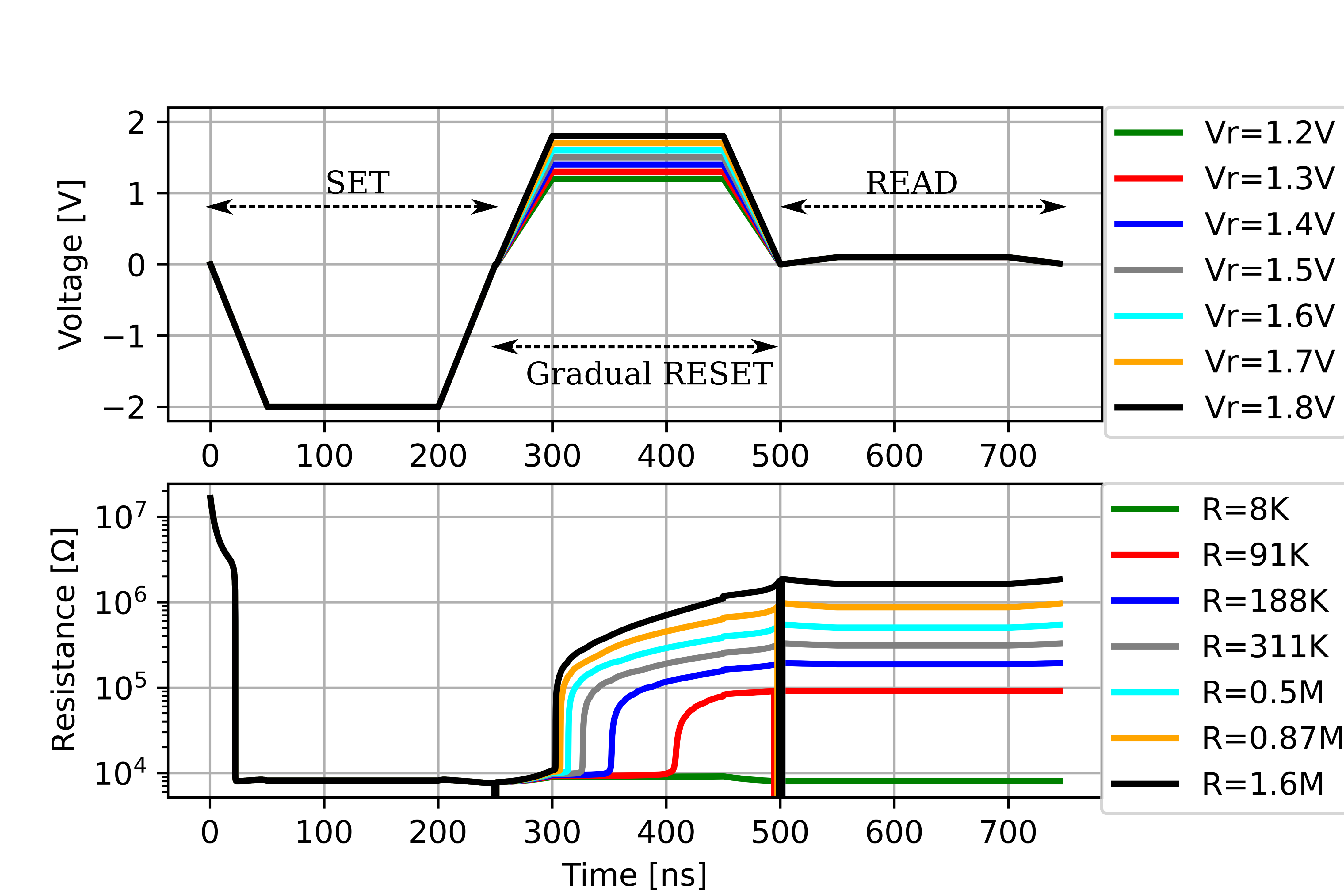}
    \caption{Multi-states behavior using gradual RESET method.}
    \label{fig:gradual_reset}
    \vspace{-1mm}
\end{figure}

\subsection{1T1R cell characterization}
\label{cell_char}
1T1R cell used for this study is designed using a Pt/Ti/TiO$_x$/HfO$_2$/Pt material stack memristive devices in series with a 45nm transistor. The material stack used in the memristive device adheres to the characteristics of the experimental devices. Fig.~\ref{fig:1T1R} shows the configuration of the 1T1R cell along with the material stack of the device and I-V characteristics. Table~\ref{tab:param} shows the parameters used for the ReRAM device model. The device has multiple-state characteristics, and the gradual RESET method has been used to achieve this behavior. In the gradual RESET method, the device is initialized to an LRS state by applying a positive voltage of a specific duration on the ohmic electrode (OE). Next, the device is switched into HRS by applying RESET pulses gradually. Fig.~\ref{fig:gradual_reset} shows the gradual RESET method, where the voltages across the device and its corresponding resistance states have been plotted. The gradual RESET method results in multi-level behavior of 1T1R cells. Additionally, a transistor in series helps to control the current precisely. 
\begin{table}[!b]
    \centering
    \caption{Model parameters}
    \label{tab:param}
    \begin{tabular}{|c|c|c|c|}
    \hline
        Symbol & Value & Symbol & Value \\
        \hline
        $\rm l_{cell}$ &  3 $nm$ & $\rm l_{det}$ & 4 $nm$ \\
        $\rm r_{det}$ &  20 $nm$ & $\rm N_{plug}$ & $20\times10^{26} \ m^{-3}$\\
        $a$ & 0.25 $nm$ & $\mu_n$&$1\times10^{-6} \ m^{2}/{Vs}$ \\ 
      $\epsilon$ &17 $\epsilon_0$ & $\rm N_{disc,min}$ & $0.008\times10^{26} \ m^{-3}$  \\ 
      $\epsilon \phi \beta $ &5.5 $\epsilon_0$ &$\rm N_{disc,max}$ & $20\times10^{26} \ m^{-3}$ \\
      $e\phi_{\beta n 0}$ &  0.3 $eV$ & $e\phi_{\beta n}$ & 0.1 $eV$ \\
      $\Delta \rm W_A$ &  0.7 $eV$ & A & 0.00392 1/$\Omega$\\
      $\rm R_{series}$ &  650 $\Omega$ & $\rm R_o$ & 719.244 $\Omega$ \\
      $\rm R_{th,line}$ &  90.47 $K\Omega$ & $\rm R_{th0}$ & $1.572\times10^7$ $\Omega$ \\
      
    \hline
    \end{tabular} 
\end{table}

\begin{figure}[!t]
    \centering
    \includegraphics[width=0.9\linewidth]{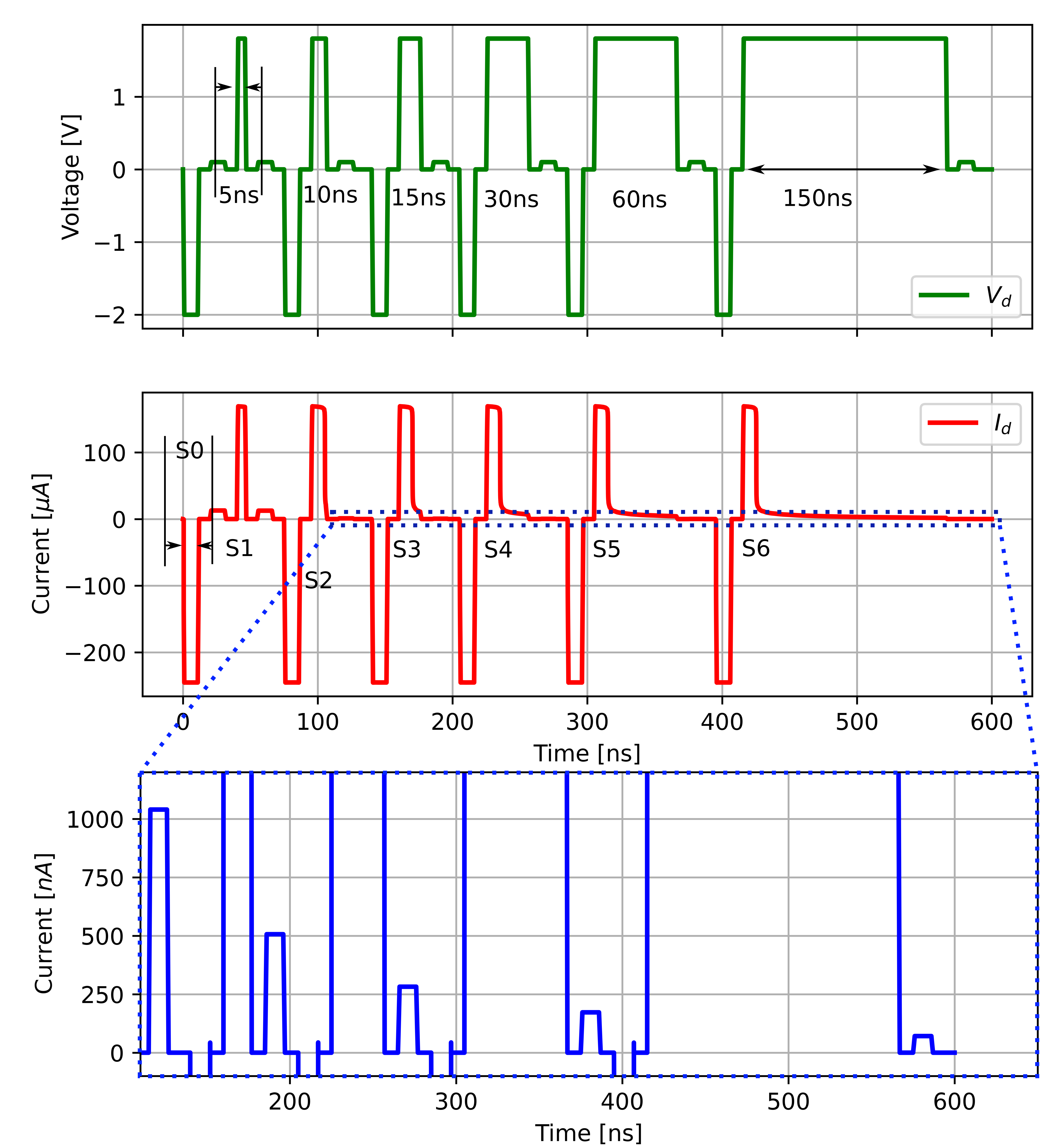}
    \caption{State switching from $S_0$ to $S_6$. For the desired state, the device is first switched to $S_0$ and then directly switched to the required state by applying the appropriate pulse. \vspace{-1mm}}
    \label{fig:states}
\end{figure}
\vspace{-1mm}
\subsection{Realization of state transitions}
\vspace{-1mm}
\label{states_trans}
 As the memristive devices change their states based on the value of integral over time of the applied voltage, the width of the applied pulse can be used for state transitions instead of gradually varying voltage. Table~\ref{tab:states} shows the switching of FA from $S_0$-$S_6$ with different pulse widths. The state $S_0$ is the initial state or LRS state, which can be switched by applying a 10ns pulse of -2V voltage. The states from $S_1$ to $S_6$ (6 states) are the actual states used for FA. Generally, the FA switches the state in the forward ($S_1$-$S_6$) or backward ($S_6$-$S_1$) directions. However, the gradual method only works in forward state switching. The backward state switching is tackled by switching to the intermediate state ($S_0$) before switching to the desired state. For example, the current state is $S{3}$, and the next expected state is $S_2$. In this case, the device first is switched to $S_{0}$ and then switched from $S_{0}$ to $S_{2}$ ($S_3\rightarrow S_0 \rightarrow S_2$). To reduce the complexity of the control circuit and accurate state detection, the intermediate state has been used in forward, and backward switching.  Fig.~\ref{fig:states} shows the switching of the states through the intermediate state. Table~\ref{tab:states} also shows the states' current and resistance, indicating enough margin between the state's current/resistance for accurate state detection. 
\vspace{-1mm}
\subsection{Impact of variations}
\vspace{-1mm}
\label{variations}
The D2D and C2C variation in ReRAM devices can affect the switching behavior. To simulate D2D variations, the random set of values for device parameters such as radius, length, and minimum and maximum oxygen vacancy in the disc are drawn from the experiment-verified Gaussian distribution~\cite{Bengel2020}. These variations were then independently applied to the available devices in the crossbar. C2C variations are simulated by changing the variable parameters in a period of a single cycle. It has been observed that states from $S_0$ to $S_3$ (low states) have a larger impact of the variations compared to high states ($S_4$ to $S_6$), which is around $\pm$50\% change in the read current for low states and $\pm$20\% for high states. However, for low states, the margin between the state is more than five times which enables the accurate detection of the state even if the variations have a larger impact. Moreover, switching through the intermediate state prevent error accumulation over time and reduces the impact of variations.
\begin{figure}[!t]
    \centering
    \includegraphics[width=0.8\linewidth]{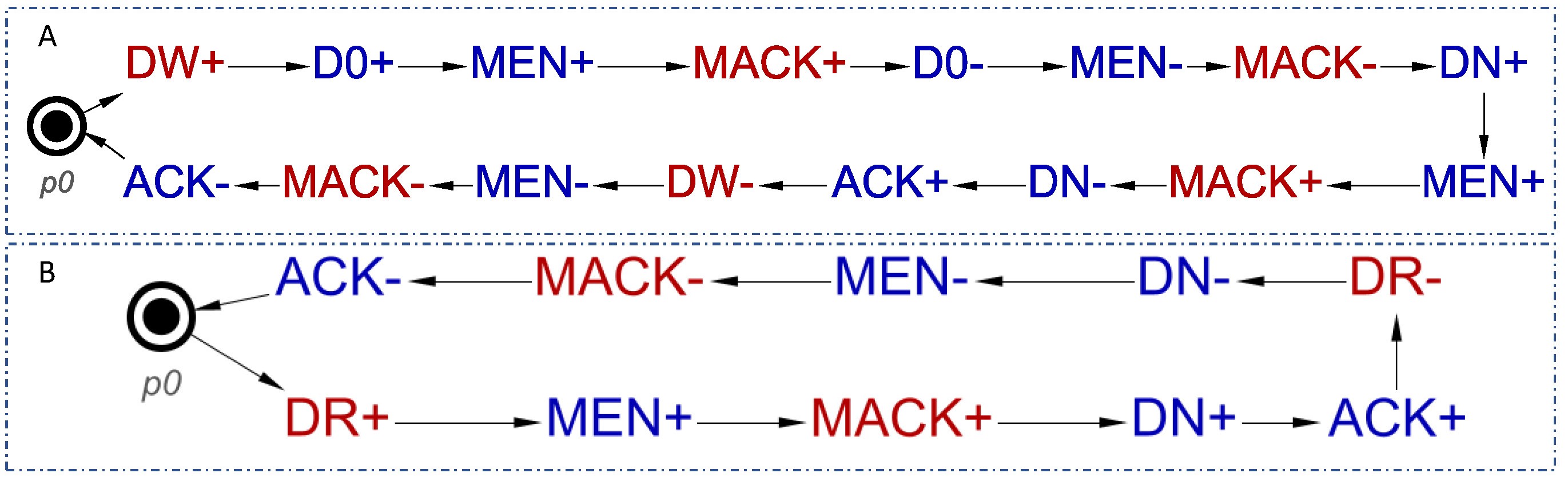}
    \caption{STG of the control unit, (a) state transition cycle, and (b)  reading the present state in FA. \vspace{-1mm}}
    \label{fig:stg}
\end{figure}

\begin{table}[!b]
    \centering
    \setlength{\tabcolsep}{2pt}
    \caption{1T1R cell energy consumption}
    \label{tab:pwr_consum}
    \begin{tabular}{|c|c|c|c|c|c|}
    \hline
         \makecell{State \\ Switching} & \makecell{ Intermediate \\ State} & \makecell{Energy\\ ($pJ$)} & \makecell{State \\ Switching} & \makecell{ Intermediate \\ State} & \makecell{Energy\\ ($pJ$)}  \\
         \hline
          $ S_{0} \rightarrow S_{1}$& -- & 1.74 & 
         $ S_{1} \rightarrow S_{2}$&$ S_{0}$ & 8.2 \\
         $ S_{2} \rightarrow S_{3}$& $ S_{0}$&8.3 &
         $ S_{3} \rightarrow S_{4}$&$ S_{0}$& 8.5 \\
         $ S_{4} \rightarrow S_{5}$&$ S_{0}$ & 8.8 &
         $ S_{5} \rightarrow S_{6}$& $ S_{0}$&9.25 \\
         \hline

        \multicolumn{3}{|c|}{\textbf{Average energy}} & \multicolumn{3}{c|}{\textbf{7.5pJ}} \\
        \hline
         
    \end{tabular}

\end{table}
\vspace{-1mm}
\subsection{Control circuitry}
\vspace{-1mm}
An asynchronous digital controller has been designed using Workcraft~\cite{workcraft-tool} to coordinate data flow between different components in the architecture. Faster operation and lower power consumption are some of the advantages of asynchronous circuits over global clocked-based circuits. The control unit's signal transition graphs (STG)~\cite{1985-yakovlev-petrinets_workshop-stg} to perform state detection (Read cycle) and state transitions (write cycle) are shown in Fig.~\ref{fig:stg}. When there is a request to read the state of FA, the control unit receives a data reading strobe (\emph{DR+}) from the digital interface environment, and a reading cycle begins. It activates MUX and bit line encoder to select a device (MEN+). Next, it starts reading the data for the $n_{th}$ FA (DN+) after receiving acknowledgment from MUX (MACK+). Lastly, the data is read, and the next read cycle is prepared by resetting DN-, MEN-, and MACK-. The final ACK signal is sent out as an acknowledgment for the digital interface.
 


The control circuit for state transitions provides a facility to transition the FA state from any present state to any other possible state. This increases the flexibility of the proposed architecture to run any FSA application. However, every state transition in FA is done via $S_0$ to maintain functional correctness and reduce controller complexity. The control circuitry handles this situation by generating an acknowledgment signal, which includes the transition of $S_0$ and desired state ($S_n$). The transition cycle is initiated whenever there is a request on a DW+ signal from the environment. The first step for state transition includes switching to the $S_0$ state. The FA is selected by enabling the row multiplexer and bit-line encoder (MEN+). The FA changes the state from the $n_{th}$ state to $S_0$ and disables the row MUX before sending the acknowledgment for the final transition. At this time of the cycle, FA is in $S_0$ and ready to be switched to the desired state ($S_n$). Similar to $S_0$ switching, the final transition starts by enabling the peripherals. Before the transition cycle is finished, the signal DN- initiates the ACK+ that will be delivered to the digital interface, resetting DW-, MEN-, and MACK-. The final acknowledgment for the digital interface is ACK- signal, which indicates a successful state transition.

\vspace{-1mm}
\subsection{Energy and latency analysis}
\vspace{-1mm}
\label{comparison}
Each state transition in FA consumes different energy, which is given in Table~\ref{tab:pwr_consum}. In every transition in FA, 7.5pJ energy is consumed on average. An FA has to switch to the intermediate state before the desired transition, which increases energy consumption. However, the intermediate state makes the state switching robust against D2D and C2C variations. As the proposed architecture used different pulse duration to state transitions, state $S_6$ takes 150ns pulse. Hence the pulse generation module generates each pulse for a 150ns period with varying widths. However, latency can be improved further by increasing the voltage amplitude of the applied pulse, which increases energy consumption. So, there is a trade-off between energy consumption and latency. \vspace{-1mm}

\vspace{-1mm}
\section{Conclusions}
\vspace{-1mm}
\label{sec:conclusions}
In this work, for the first time, we proposed the architecture to implement the FSA using a 1T1R ReRAM crossbar. This paper offers insights into the scope of FSA utilizing ReRAM and CMOS technology. We use the multi-level characteristics of ReRAM, achieved using the gradual RESET method, to implement FSA on the crossbar. We studied the impact of variation on state transitions. Finally, we evaluated the proposed framework in terms of latency and energy consumption. The results are encouraging and demonstrate the potential for using ReRAM-based FSA designs. We will explore the prototyping of the proposed designs and test the architecture with learning automaton applications in the future.


\vspace{-1.5mm}
\section*{Acknowledgments}
\vspace{-1mm}
This work was supported in part by the Federal Ministry of Education and Research (BMBF, Germany) in the project NEUROTEC II under Project 16ME0398K, Project 16ME0399, German Research Foundation (DFG) within the Project PLiM (DR 287/35-1, DR 287/35-2) and through Dr. Suhas Pai Donation Fund at IIT~Bombay.
\balance
\bibliographystyle{IEEEtran}
\bibliography{Bib}

\end{document}